\documentclass[12pt]{article}
\usepackage{epsfig}
\usepackage{psfrag}
\usepackage{amsmath}
\usepackage{lscape}
\usepackage{multirow}
\usepackage[super, comma, sort&compress]{natbib}
\newcommand{\bfig}{\begin{figure}[h] \begin{center}}
\newcommand{\efig}{\end{center} \end{figure}}

\begin{document}

\begin{center}
{\large Biomolecular transitions: efficient computation of pathways, 
free energies, and rates}
\end{center}
\begin{center}
Divesh Bhatt and Ivet Bahar
\end{center}

\section*{Abstract}

 We present an efficient method to compute transition rates
between states for a two--state system. The method utilizes
the equivalence between steady--state flux and mean first passage
rate for such systems. More specifically, the procedure divides the
configurational space into smaller regions and equilibrates
trajectories within each region efficiently. The equilibrated
conditional probabilities between each pair of regions lead to
transition rates between the two states. We apply the procedure to
a non--trivial coarse--grained model of a 70 residue section of
the calcium binding protein, calmodulin. The procedure yields a
significant increase in efficiency compared to brute--force
simulations, and this efficiency increases dramatically with
a decrease in temperature.

\section{Introduction}

 Biomolecular processes frequently involve conformational transitions.
Some examples of such biomolecular transitions occur during substrate
binding, protein folding, and motion of molecular motors. An
understanding of the pathways for these transitions is crucial
in determining how these processes work, and what are the possible
intermediates along the pathways. For example, there is
a significant focus on understanding the protein--ligand binding
process in order to determine the extent to which the presence of ligands
influence the conformation change in protein.\cite{hammes-oas,Weikl2009}

 In a significant number of studies, transitions between two states
are of interest. Transitions between a unique folded
state and unfolded ``state'' of a protein, and conformational transitions
between the apo and holo conformations of an enzyme are examples
of such two--state problems. Indeed, experimental transition rates
are also reported between two states (for example,
transitions between open and closed conformations of adenylate
kinase\cite{adk_chu}). In this manuscript, we focus on presenting a methodology
for studying the transitions and computing the associated rates
efficiently in such two--state problems.

 Studying biomolecular transitions in atomistic detail via 
computational techniques is difficult due to timescales
involved ($>$ ms, typically). Several techniques are used for
circumventing such large timescales for biomolecular
transitions. These range from approximate methods such as
targeted and steered molecular dynamics,\cite{tmd1,tmd2,tmd3}
to rigorous nudged elastic band (NEB)\cite{neb3,neb4,neb5,neb6}
and related methods.\cite{action1,action2,action3,action4,Pan08}
The use of an extra driving force in targeted and steered MD precludes
the determination of correct transition rates for the
unperturbed system, whereas the above noted rigorous methods focus on
the determination of a minimum energy path for transition and
not the transition rates.

Coarse graining is another approach to study transitions. Coarse--grained
network models have been used to study conformational transitions
efficiently,\cite{Temiz04,adk_cg1,adk_cg2,adk_cg3,adk_cg4,Yang09}
however, it is not possible to obtain transition rates consistent
with the particular network model directly. Brute--force and
path--sampling simulations have also been used with coarse--grained
models to determine transition rates and pathways.
These methods give rigorous path ensembles and
transition rates, provided that statistical sampling of the
relevant configuration space is possible.

 The goal of this manuscript is to present an efficient method of
computing statistically rigorous transition rates (and
accurate transition paths) for a given forcefield, 
given approximate transition paths between two states. 
The initial transition paths may be generated via faster coarse--grained
models and/or elevated temperatures (and, hence, approximate paths for the 
desired model and temperature).
Subsequently, the regions around structures along the transition paths
are equilibrated and inter-bin transition fluxes are computed, leading
to transition rates between the two end states. 

 The manuscript is organized as follows. First, we discuss in detail
the procedure and highlight the underlying concepts. Then, we
describe the non-trivial system we simulate as a test case. This is
followed by detailed results. Next, we discuss the efficiency of
the procedure. Finally, we give the conclusions obtained.

\section{Method}

 We first discuss the underlying ideas behind our approach, followed by
details on the implementation.

The crucial point that we exploit in our approach is the
equivalence between steady state flux from one state to another
(say, from state A to state B) and the mean first passage rate (or,
inverse mean first passage time, MFPT).
from state A to state B, provided that a trajectory that enters
state B is immediately fed back into state A.\cite{hill2} 
This equivalence is given by
\begin{equation}
\mathrm{SS_{flux} (A\rightarrow B)}=\frac{1}{\mathrm{MFPT (A\rightarrow B)}}.
\label{e2}
\end{equation}
We compute the flux from one state to another while maintaining
the states at equilibrium (equilibrium is a special steady state).
Thus, our approach relies on establish of such an equilibrium, and the
calculation of fluxes.

 We begin by realizing that a small region of the relevant configuration 
space is faster to equilibrate than the whole (relevant) configuration space. 
Thus, the main procedural idea is to equilibrate ``small'' regions of 
the configuration space and compute transition probabilities from such 
a region to other regions. If the trajectories in a region of configuration
space are distributed according to the equilibrium distribution, the
transition probabilities out of that region into neighboring regions
are stationary. This self--evident property follows from the
stationarity of the equilibrium distribution itself. 

 The ratio of transition probabilities between two well--equilibrated
regions gives the ratio of equilibrium populations of the two regions.
In other words, if $p (j,\Delta t|i,0)$ represents the transition
probability from state $i$ to state $j$ in time $\Delta t$,
\begin{equation}
\frac{N_i}{N_j}=\frac{p (j,\Delta t|i,0)}{p (i,\Delta t|j,0)},\quad
\mathrm 0<\Delta t<\infty
\label{e1}
\end{equation}
where $N_i$ is the equilibrium population of state $i$. The equality
is valid for all values of $\Delta t$, since the (statistically averaged) 
probability fluxes between two states must balance each other exactly for
observations at all time increments (equilibrium state is unperturbed).
Figure~\ref{f0} show a schematic
of a two-dimensional system. Once the equilibrium probability
distribution within a region is reached (irrespective of other regions),
the transition probability from that region is stationary.
Because we use eq~\ref{e2},
$\Delta t \ll {\mathrm MFPT}$ to approximate {\em immediate} feedback.

 Thus, the procedure is as follows: divide the relevant configuration space
into smaller regions, establish equilibrium within such a region and compute
transition probabilities between regions, compute populations of
these regions and combine them into two final states, and calculate
transition rates (via equilibrium fluxes) between the two states. We 
discuss each of these steps in greater detail below.

\subsection{Division of relevant configuration space}

 First, we discuss how the relevant configuration space is divided
into regions. For transition between different protein conformations (or
states) the relevant part of the configuration space lie along the
transition path(s) between the two states. Similarly, folding
pathway(s) define the relevant part for equivalent problems in folding.

 For a large variety of models, the generation of these transition
pathways itself is a challenge. To address this issue, coarse--grained
models or elevated temperatures can be used to generate approximate 
transition pathways and the regions can be defined by Voronoi constructions 
around the structures along these pathways. Among coarse--grained models, 
network models have shown
significant success in determining these transition paths quickly,
and are good candidates for identifying structures along the
transition pathways. 

 Subsequently, simulations using the  model of interest can be started 
from these identified structures to
equilibrate trajectories in the Voronoi regions around these
structures.

\subsection{Equilibrium within a region}

We desire equilibration within
a region and achieve this using the above mentioned property of the
system at equilibrium: a system once established at equilibrium 
will remain at equilibrium.
Thus, if a trajectory leaves a region from a point and is immediately
fed back into that same point in that region, equilibrium will be maintained. 
As $\Delta t\rightarrow 0$, this procedure corresponds to marking the
exit points on the surface of a region and using these points as
entry points for trajectories back into the region.

\subsubsection{Efficient attainment of equilibrium within a region}
\label{eff}

 The procedure mentioned above establishes equilibrium
within a region after the molecular relaxation time within the region.
However, this relaxation time itself is often substantial. Thus,
in this section, we discuss additional steps that ``enhance''
the establishment of equilibration within the region.

 Firstly, we realize that the ``brute--force'' approach to establishing
equilibrium within a region is to originate
a multitude of trajectories with equal weights from the reference 
structure for the Voronoi region.
In contrast, we utilize an adaptive procedure, that
is based on enhanced steady--state path sampling.\cite{ss} In particular,
transition probabilities between bins (plus the requirement of
total probability being unity) can be used to estimate
populations within each bin (as shown in eq~\ref{e1}, except, now, for bins
within a region).

We start $N$ trajectories with equal weights ($=1/N$) from the
reference structure, propagate the trajectories for a short
simulation time, feed any trajectory that escapes the Voronoi 
region back into the region, as discussed above,
and calculate the distribution of trajectories
as a function of distance from the reference structure.
Then, we compute the mean and variance of this distribution, and accordingly
divide the region into $M$ bins. Subsequently, we perform
combining and splitting of trajectories as done in the weighted ensemble
method to obtain
$N/M$ trajectories in each bin 
(with, possibly, different weights).\cite{huber,bin}
The subsequent steps of the procedure are enumerated below.

\noindent
{1.} Propagated each trajectory for a time $\Delta t$, and compute the
new distribution of the trajectories (after feeding back any trajectories
that left the Voronoi region).\\
{2.} Compute transition probabilities between bins within a Voronoi
region, estimate populations in each bin, and reassign weights to 
trajectories in each bin based on the population estimate in that bin.\\
{3.} Recompute mean and variance of the distribution of trajectories 
as a function of distance from the reference structure for that
Voronoi region, and recompute new $M$ bins.\\
{4.} Split and combine trajectories within a bin such that each bin
has $N/M$ trajectories.

 Steps 1--4 are repeated until a criteria for convergence, discussed
below, is satisfied.

\subsubsection{Convergence within a Voronoi region}
\label{dkl}

 Upon equilibration within a Voronoi bin, all transition
probabilities from that bin attain stationary values. Thus,
a natural test for convergence within a bin is when the distribution of
transition probabilities from a bin stops evolving. In particular,
we check Kullback--Leibler (KL) divergence\cite{kl} of block-averaged
distributions of previous blocks from the current block. KL
divergence between a ``true'' distribution $p_{ij}$ (e.g., transition
probability from state $i$ to state $j$), and an approximation $q_{ij}$
is defined by
\begin{equation}
D_i^{\mathrm{KL}}=\sum_{j}p_{ij}\ln\frac{p_{ij}}{q_{ij}}.
\label{e3}
\end{equation}

 In the current context, we take the true distribution as
the average in the current block, and $q_{ij}$ as transition
probabilities in previous blocks. If several of the previous
block averages differ by a small value from the current block
average, convergence is assumed, and the stationary
transition probabilities {\it between different Voronoi regions} are
subsequently accumulated for averaging.

\subsection{Transition probabilities between regions}

 Transition probabilities out of an equilibrated region are computed
by monitoring the conditional probabilities, $p(j,\Delta t|i,0)$.
The time increment $\Delta t$ is arbitrary in the sense that
the populations (cf. eq~\ref{e1}) are independent of it. However, for a large
value of $\Delta t$, $p(j,\Delta t|i,0)\sim N_j$ (the equilibrium
distribution itself), and
there is no dynamical information present at this extreme.

 In essence, $\Delta t$ is the time interval at which an observation
as to whether the system is in state $j$ is made. And the chosen value
represents the ``granularity'' of time. We reiterate that eq~\ref{e1} is valid
at equilibrium for any value of $\Delta t$ (however, the use of
eq~\ref{e2} to calculate subsequent rates dictates that 
$\Delta t \ll {\mathrm MFPT}$).

\subsection{Populations of regions}
\label{pop}

 The equilibrium probability distribution in each region can be
computed from the transition probabilities (as shown in eq~\ref{e1},
followed by subsequent normalization).
Repeated multiplication of an arbitrary starting probability
distribution in the Voronoi regions by the stochastic transition
matrix leads to a stationary distribution of populations in
each Voronoi region.

 There is variability associated with the transition probabilities
generated from finite--length simulations, and this variability
must be mirrored in the stochastic matrix. We generate
averages of the transition rates for $B$ blocks. If the relevant
space is divided into $R$ regions and each of the $B$ blocks forms
a unique stochastic transition matrix, the total number of
transition matrices is $B^R$.

 In principle, populations of the regions must be calculated for
each of these $B^R$ matrices individually, giving rise to $B^R$ sets
of populations. For the current application, we choose $B=4$ and $R=36$,
making it practically impossible to do such an explicit computation.
Instead, we perform sequential multiplications by randomly selected
transition matrices from the set to estimate populations of regions.
As we show below, this leads to a robust convergence of population
estimates after a relatively small number of such multiplication steps.

\subsection{Rate computation between states}

 In this section, we elaborate on the computation of rates between
states associated with biomolecular transitions. We
address the following issues: the determination of appropriate states
using the Voronoi regions, and the determination of transition rates between
those states. 

\subsubsection{Determination of the two states}
\label{comb}

We combine the Voronoi regions such that there are two states left after the
combination. In particular, we perform a simulated--annealing Monte Carlo
to combine regions such that the two obtained states
lead to the minimization of flux between the states. This corresponds to
determining the highest free energy barrier that divides the
configurational space into two states.

 The Monte Carlo sampling proceeds as follows. First, a state is
chosen at random, followed by random selection of a region (except for
the two end regions that are permanently assigned the states) in that
state. The state that the region belongs to is attempted to be switched.
The switch is successful if it leads to decrease in the overall
flux between the two states. Otherwise, the switch is accepted with
a probability given by $exp(-\alpha\Delta f)$ (where $\Delta f$ is the
change in the flux, and $\alpha$ is a scaling constant). The
assignments of regions to states that lead to the minimum in flux via
this simulated annealing procedure defines the two states.

\subsubsection{Determination of transition rates between states}
\label{rates}

Our procedure explicitly establishes equilibrium (both probabilities
and fluxes) between states. To obtain steady--state flux from state A
to state B (corresponding to a feedback as required by eq~\ref{e2}), 
we divide the equilibrium flux from state A to state B by the
population of state A.  This is appropriate because a system at 
equilibrium can always be decomposed into two steady states with
such feedbacks.\cite{bhatt11}
Equation~\ref{e2}, then, gives the transition rates (1/MFPT) from
state A to state B.

 Similar to the computation of state populations from $B^R$ possible
instances of the transition matrix, we compute equilibrium fluxes
between the two states via random selections of the instances
of the transition matrix.

\section{System and simulation details}

 As a proof of principle, we use a
a coarse--grained model for calmodulin and study the equilibrium
populations of regions along the transition paths and the rates between
its two states (PDB id: 1CLL and 1CFD).

 We use a double-Go alpha--carbon 
model,\cite{dmz_dgo,hummer_dgo,adk_cg4,onu_dgo}
and propagate the system via a 
Cartesian Monte Carlo (MC) simulation: at each step, a random residue is
selected and a displacement move is attempted (and accepted or
rejected based on the Metropolis criterion\cite{Metropolis}). Further, we
do not use any additional force to facilitate the transition --
the system evolves via natural dynamics as described by MC.

 Intermediate structures along the transition path are identified
from an actual transition trajectory obtained by running a short
simulation (note: this trajectory can also be obtained by using a
fast method such as ANM) at an elevated temperature. 
In all, we use a total of 34 intermediate
structures to construct Voronoi bins around them. Subsequently, 1000
trajectories started from these intermediate structures, and
the enhanced equilibrium attainment procedure, described above, is
used to equilibrate the system. 
Further, the
new Voronoi bin of each trajectory after 10000 MC steps is noted 
($\Delta t=10000$ MC steps).
If a trajectory has left its original region after 10000 MC steps, it is
fed back in the exact same location from which it was started in 
that region (as noted above, this allows the region to relax to
equilibrium). The process is repeated till the transition probabilities
to other regions reaches a (statistical) stationary value, as
determined by eq~\ref{e3}, indicating that
the equilibrium distribution if reached within that region.

 The stationary transition probabilities are, then, used to calculate the
fractional populations of all the regions via matrix multiplication for
50,000 steps. The state definitions are determined using results for $T=0.45$
(the same state definition is used for both temperatures), using
MC simulated annealing for $10^{6}$ steps. Finally, equilibrium
fluxes between the states are computed for $10^6$ random instances
of the transition matrix.

\section{Results}

\subsection{Equilibration within each Voronoi region}

 The crucial requirement is the equilibration of trajectories
within a region. Thus, we first discuss this aspect. In
particular, we look at the distribution of trajectories within
each region (trajectories in each region are initialized at its 
reference structure).

 Figure~\ref{f2} shows the convergence of trajectories for two
of the regions (1CFD and an intermediate). Clearly, convergence is
reached at both the temperatures with increasing time increment, $\tau$.
In all the cases, the trajectories within a region spread a substantial 
distance away from each reference structure, with the effect being
more pronounced at higher temperature. Further, the spread of
trajectories is greater at the higher temperature.

 One important point that emerges from Figure~\ref{f2} is that
time required for equilibration within a bin is similar at
the two temperatures. This is due to the use of probability
adjustment procedure as described in Section~\ref{eff}: without
the use of such a procedure, it takes significantly more
time to equilibrate within a region at a lower temperature (data
not shown). This observation of similar equilibration times at the
two temperatures has a profound significance: the computational times to obtain
steady state rates between two states at the two different temperatures
are similar. This is in sharp contrast to brute-force sampling to
determine steady state rates.

 Beyond equilibration within a region, we explicitly seek a convergence
of transition probabilities between each pair of regions. A measure
for this convergence is discussed above in Section~\ref{dkl}.
Figure~\ref{f3} shows $D_{\mathrm{KL}}$ for the trajectories in
the 1CFD Voronoi bin for three different instances of the final
block averaged values of the transition rates. Clearly, as
$D_{\mathrm{KL}}$ is computed after longer times, more and more
previous block averages are similar ($D_{\mathrm{KL}}<10^{-2}$)
to the corresponding final block averages. For the 1CFD region
shown in Figure~\ref{f3}, four previous block averages
give transition rates with $D_{\mathrm{KL}}<10^{-2}$ for the green
symbols. ``Production'' transition probabilities are, then,
computed for further 60 $\tau$.

 The convergence of transition probabilities is robust with
regards to block size: when blocks of $20\tau$ are used,
identical results are obtained.

\subsection{Transition probabilities}

 Equilibrated transition probabilities are key towards computing
populations of regions and transition rates between states.
 The final averaged transition probabilities between regions are shown 
in Figure~\ref{f4} for the two temperatures. Two distinct blocks
of regions emerge from each panel of Figure~\ref{f4}, indicating
a clear two-state decomposition of the sampled conformational space.
Further, as expected, the transition probabilities are more diffuse at the
higher temperature.

 Using these transition probabilities, we, then, compute the population
of each region as described in Section~\ref{pop}, followed by a
combination of regions into a total of two states.

\subsection{Populations of regions and combination into states}

 Successive multiplications of an arbitrary initial distribution of
populations by a stochastic matrix lead to (rapidly) converged
estimates of populations of the regions. Since
we have a distribution of stochastic matrices (due to estimated
errors in the transition probabilities), we obtain fluctuating values of
region populations, as shown in Figure~\ref{f5} for two of the
regions. However, since the different estimates of the stochastic matrix
are similar (reasonable statistical precision and converged
transition probabilities), the distribution of fractional populations
around the mean is fairly narrow.

 Figure~\ref{f6} shows the averaged populations of different regions
for both the temperatures. Clearly, the two end states (regions 1 and 36,
corresponding to 1CLL and 1CFD, respectively) show the high populations,
whereas intermediate regions far from both 1CLL and 1CFD show low
populations. A combination of regions into states, as discussed in
Section~\ref{comb}, leads to the identification of each region with
one of the two states (as shown by different symbols, circles and
triangles, in Figure~\ref{f6}).

 As temperature increases, we expect fractional populations to
become more diffuse: population from basins is redistributed into
intermediate structures. Clearly, a comparison of populations of each region at 
the two temperatures in Figure~\ref{f6} shows that the two end regions 
show a higher population at the lower temperature, whereas for intermediate 
regions, the population is higher at the higher temperature.

 Another feature that emerges from Figure~\ref{f6} is the apparent
``roughness'' of the free energy (negative of the ordinate in Figure~\ref{f6}) 
surface in the 1CFD state as compared to the 1CLL state. A related feature is
that the regions in the two states that are closest to the transition
region(s) between the two states show higher populations than regions
further inside a state (compare region 16 with region 15, for example),
suggesting that the free energy surface shows an apparent local
minimum near the transition. However, it is not possible to assert
these observations conclusively since the volumes of the Voronoi
regions may not be equal.

 A summation of populations of regions gives the state populations.
At $T=0.45$, the 1CLL state has a fractional population of 0.34. A
Similar value (0.35) is obtained at $T=0.4$.

\subsection{Transition rates between states}

Populations of regions, combination of
regions into states, and transition probabilities lead to the
computation of transition flux between the two states 
(cf. Section~\label{rates}). Figure~\ref{f7} shows the
transition fluxes between the two states (symmetric) for the two
temperatures. Again, we obtain a distribution of fluxes that are
fairly close to the mean values. Further, as expected, the fluxes
between the two states are higher at $T=0.45$.

 The computed transition rates between the two states are
reported in Table~\ref{t1}. Decreasing the temperature from
$0.45$ to $0.4$ decreased the transition rate by approximately
a factor of 2 in either direction.

\section{Discussion}

\subsection{Efficiency compared to brute--force simulations}

 The main aim of this manuscript is to present an efficient method
to compute transition rates for a two--state system. Thus, in this
section, we compare the time required for computing transition
rates via this method with an estimate of time required via
brute--force simulations.

 The total number of Monte Carlo steps required comprise of mainly
the time required to establish equilibrium and the time for
production runs for transition probabilities in each region (the
generation of initial path(s) at elevated temperatures, or via
fast network models is a small fraction of this time, and the
time for matrix multiplications for computing populations
of regions and, subsequently, rates is negligible). Since we
use a thousand trajectories in each region, the total number
of MC steps are approximately $5\times 10^{10}$, at {\it either}
temperature.

For establishing equilibrium via brute force throughout the configurational
space, there must be several traversals of a trajectory between the
two structures. The
 transition rates (based on MC steps) from the 1CLL structure to 
the vicinity of 1CFD
structure were computed by Zhang {\it et al.}\cite{bin} at $T=0.45$ and
$T=0.4$ to be $7\times 10^{-11}$ and $8\times 10^{-12}$, respectively.
(We emphasize that these transition rates are from one structure to
the vicinity of another structure and {\em not} from one state to
the other: they cannot be compared directly). The mean first passage times
from 1CLL to the vicinity of 1CFD are, thus, $1.4\times 10^{10}$ and
$1.3\times 10^{11}$ MC steps at $T=0.45$ and $T=0.4$, respectively.
Clearly, if several such passages are required in {\em both} directions,
brute--force simulations are significantly less efficient
than the method discussed in this manuscript. Further, the efficiency of
brute-force simulations is dramatically decreased at lower temperatures,
in contrast to the method presented.

\subsection{Contrast with Markov models and applicability}

 The method presented here bears some similarities to 
Markov models\cite{Ozkan02,chodera}
that are used to study transitions between states. However,
there is an important distinction. In Markovian analysis, several
microstates (similar to regions in the present work) are combined,
based on rates between them, to form macrostates. Subsequently,
the rates between the macrostates can be computed based on rates
(or, transition probabilities) between the microstates. The combination
into {\em Markovian} macrostates allows one to consider only the rates between
states, and not the equilibration within a state.

 In contrast, the use of the equivalence between steady--state flux and
the mean first passage rate in the current work allows for the
relaxation of the Markovian condition.
Instead, the 
computed rates by the current method are steady--state
rates for the situation when steady state is established by feeding the
trajectories that enter into the target state {\it immediately} back
into the starting state (and the feedback is such that the equilibrium
distribution in the initial state is maintained). Thus, in effect,
the computed rates are equilibrium rates between the two states.

 Two aspects emerge from the preceding discussion - both due to the 
equivalence of steady--state flux and mean first passage rate. First,
the feedback must be immediate for this relation to be exactly valid. 
In contrast, $\tau\equiv\Delta t=10000$ MC steps above. However, if the 
two states are
separated by a barrier, any trajectories that enter a different state
{\em between} two updates, separated by $\tau$, are unlikely to escape the new
state during a $\tau$ interval. Thus, we expect that relation to be
valid for range of $\tau$. We emphasize that the
presence of such a barrier does not necessarily imply that either state
behaves in a Markovian fashion within the $\tau$ interval: the
time for equilibration within either state may be significantly more than
$\tau$ without affecting the validity of the current approach.

 The second aspect is that the explicit use of the equivalence
restricts the use of the approach to two--state problems. However, as
mentioned in the Introduction, such two--state problems are
encountered very frequently.

\section{Conclusions}

 In this manuscript, we present a method to compute transition rates
between states for two--state systems that are frequently encountered
in enzyme kinetics and protein--folding studies. The crucial idea
we employ is the equivalence between steady--state flux and
the mean first passage transition rate, and that the
division of the configurational space into smaller regions allows for
a faster equilibration within a region. We find that an adaptive
enhancement procedure greatly facilitates equilibration within a
region, allowing for an efficient computation of conditional
transition probabilities between regions. Further, this adaptive
enhancement procedure is not substantially affected by a decrease
in temperature, making the procedure increasingly efficient (when
compared to brute--force simulations) as the temperature is
decreased. This leads to an efficient calculation of state populations
and transition rates even as the temperature is decreased.

\clearpage

\bfig
\includegraphics{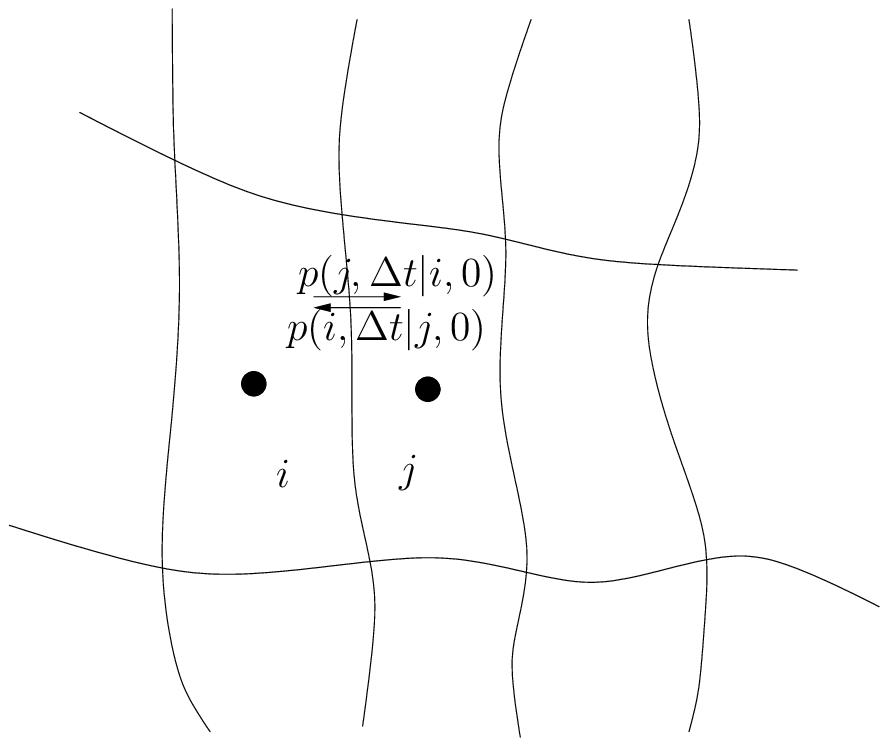}
\caption{A schematic of division of the relevant configuration space
into Voronoi regions ({\it e.g.}, any point inside region $i$ is closer to
the structure defining region $i$ (represented by a solid circle) than to
the structure defining any other region. Two such regions are explicitly
labeled with transition probabilities between them depicted.}
\label{f0}
\efig

\clearpage

\bfig
\includegraphics{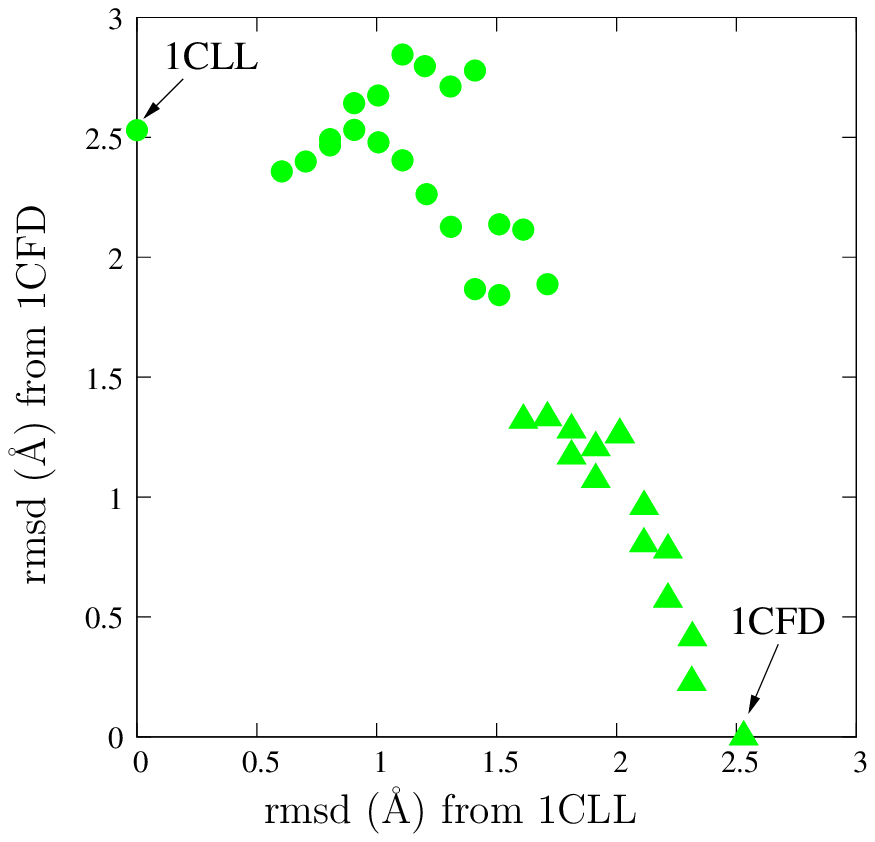}
\caption{Structures used to construct Voronoi bins along a putative 
transition pathway between the
two crystal structures of calmodulin (1CLL and 1CFD) plotted as
rmsd from the two structures. The circles represent structures that
are closer to 1CLL, whereas the triangles represent structures closer to
1CFD. The two known crystal structures (1CFD and 1CLL) are also shown
labeled.}
\label{f1}
\efig

\clearpage

\bfig
\includegraphics{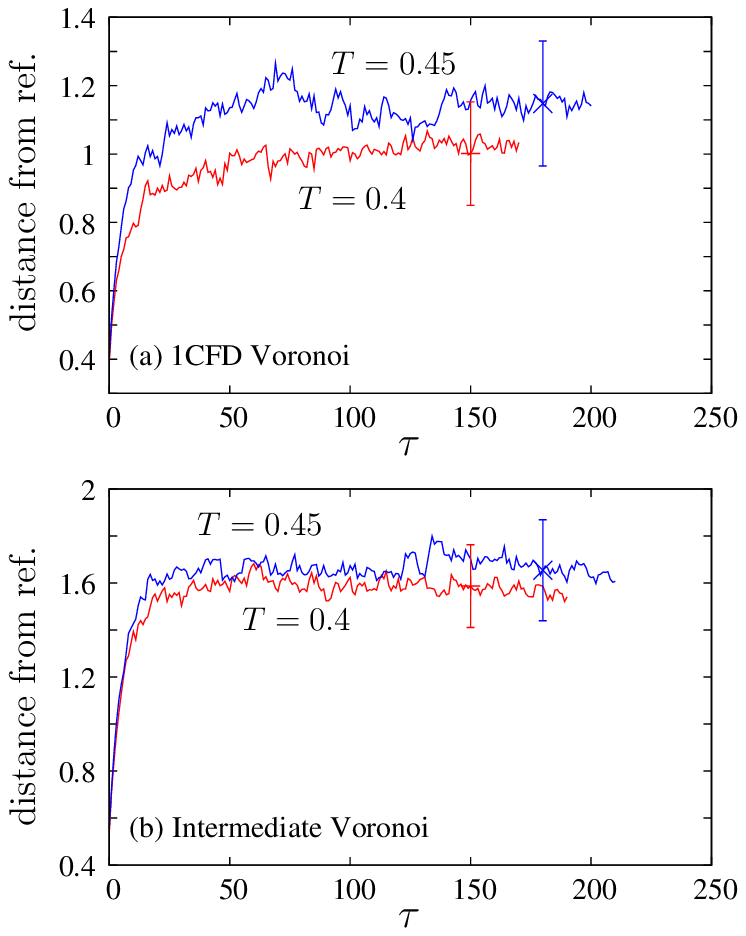}
\caption{Convergence of trajectories in Voronoi bins as functions of
time increment $\tau$. Panel (a) shows, for two different temperatures,
the distance of trajectories in the 1CFD bin from the 1CFD structure.
Panel (b) shows similar results for an intermediate structure. The spread
associated with the trajectories in a bin is also depicted at one
specific $\tau$. ($\tau=10000$ MC steps).}
\label{f2}
\efig

\clearpage

\bfig
\includegraphics{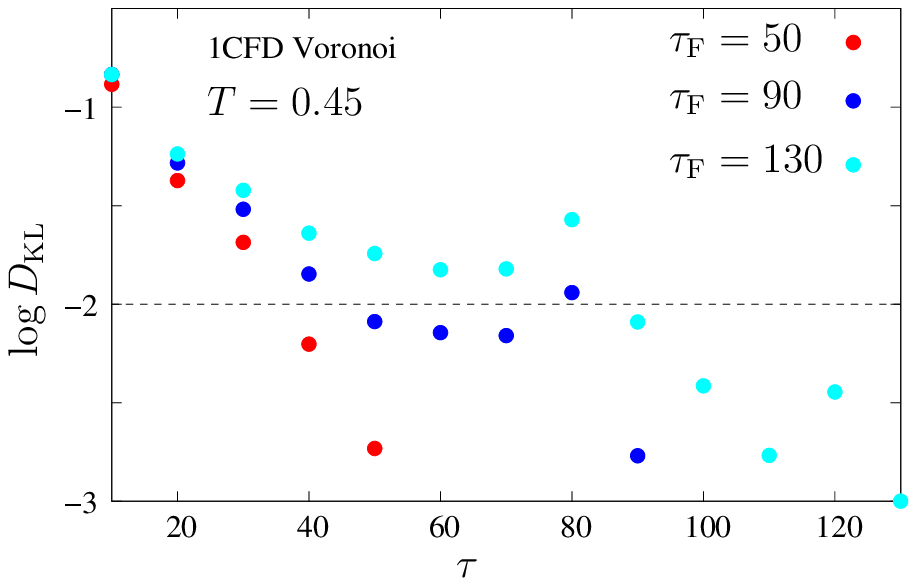}
\caption{Convergence of transition rates from the 1CFD Voronoi region to
other regions as a function of time. KL divergence (eq~\ref{e3}) from
the final block (averaged over 10 $\tau$) is compared with the
KL divergence from previous blocks. The figure shows such a comparison for
three different instances of the final block ($\tau_{\mathrm{F}}$). The
reference line for $D_{\mathrm{KL}}=10^{-2}$ is also shown.}
\label{f3}
\efig

\clearpage

\bfig
\includegraphics{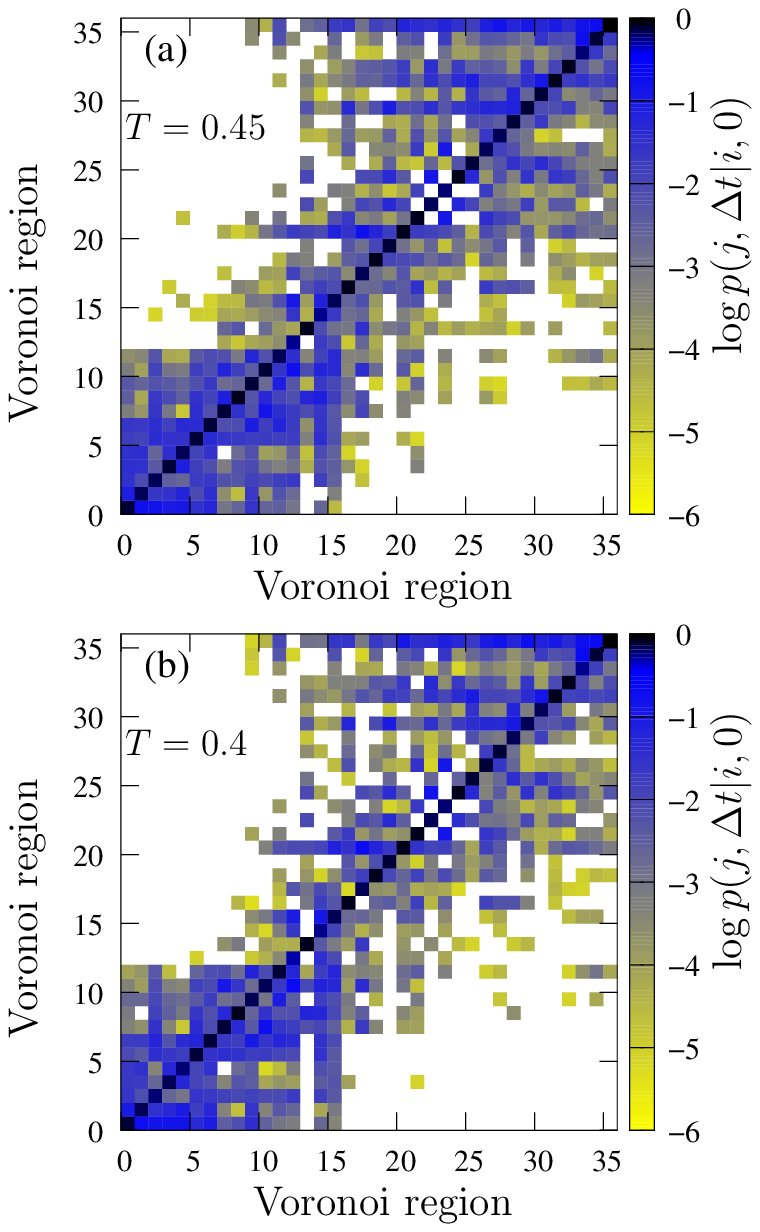}
\caption{Transition probabilities between each pair of regions -- (a)
$T=0.45$, and (b) $T=0.4$, for $1\tau = 10000$ MC steps.}
\label{f4}
\efig

\clearpage

\bfig
\includegraphics{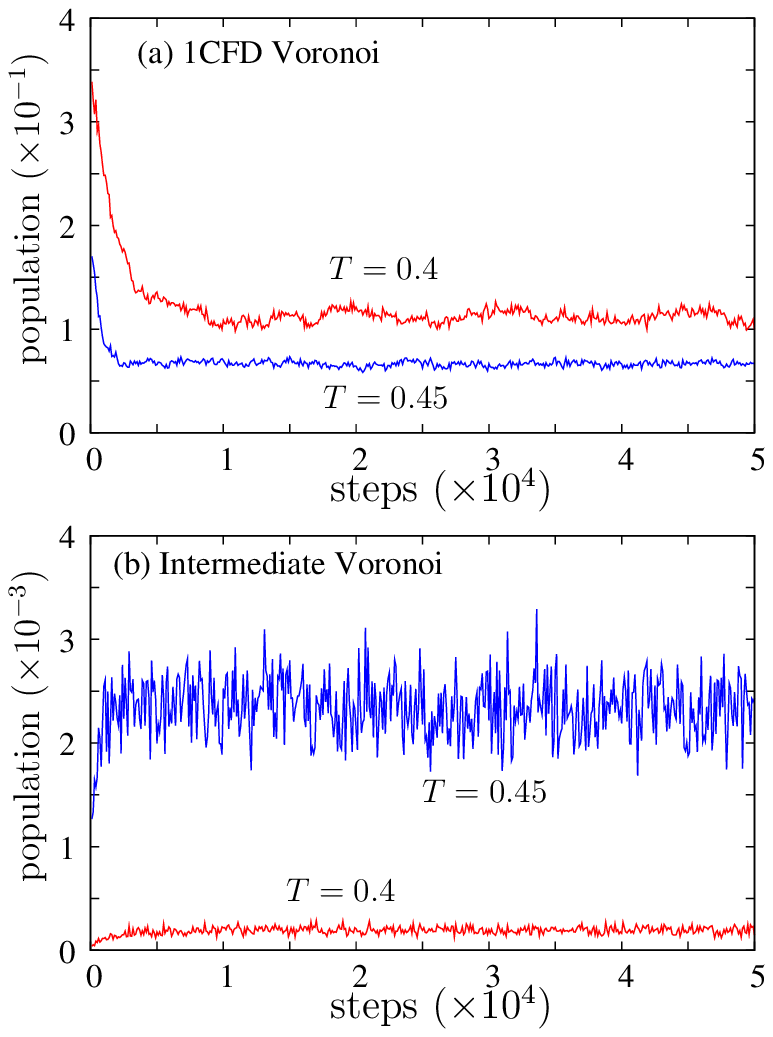}
\caption{Rapid convergence, at $T=$ 0.45 and 0.4, of fractional population 
in each state (shown for (a) 1CFD Voronoi region, and (b) an intermediate 
Voronoi region) upon sequential application of randomly selected 
stochastic matrices.}
\label{f5}
\efig

\clearpage

\bfig
\includegraphics{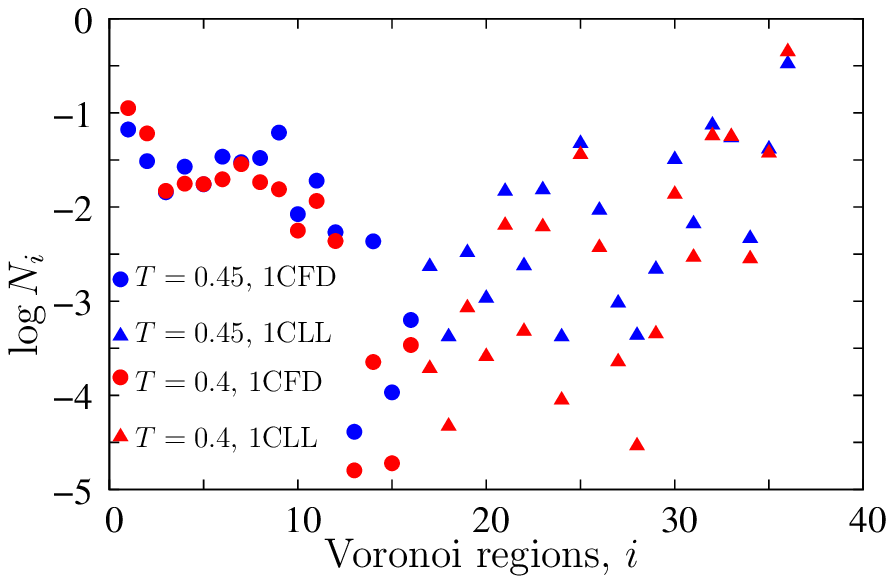}
\caption{Comparison of fractional populations of different Voronoi regions
at two different temperatures. The two extreme regions, $i=1$ and $i=36$
represent 1CFD and 1CLL regions, respectively. Other regions 
(shown by circles and triangles) belonging to
either of the two states (1CFD and 1CLL {\it states}) are shown
by identical symbols.}
\label{f6}
\efig

\clearpage

\bfig
\includegraphics{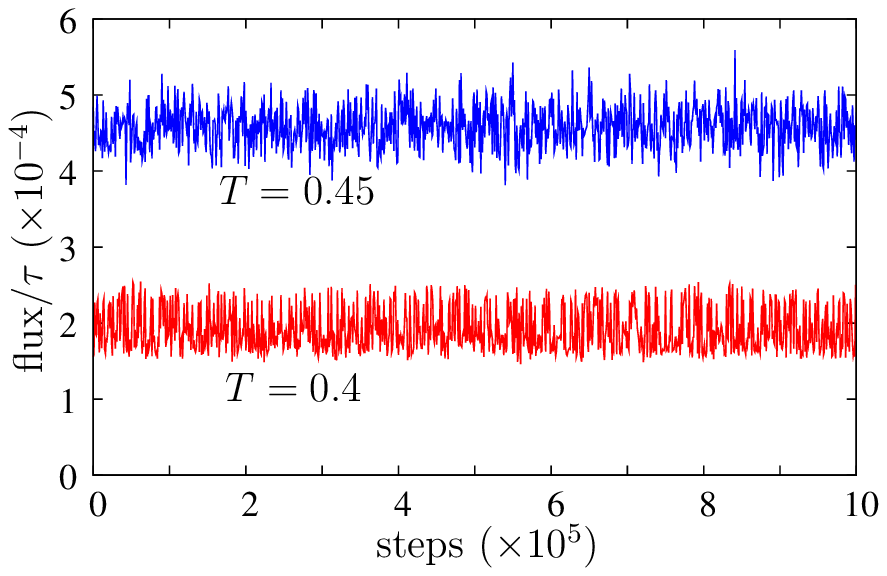}
\caption{Comparison of fluxes between 1CFD and 1CLL states as a function
of sequential random selection of the stochastic transition matrices. The
distribution of flux values gives an indication of the confidence
region for each flux.}
\label{f7}
\efig

\clearpage

\begin{table}[h]
\begin{center}
\caption{Transition rates between the two states.}
\begin{tabular}{c|c|c}
$T$ & 1CLL to 1CFD (/MC step)  & 1CFD to 1CLL (/MC step) \\ \hline
0.45 & $1.3\times 10^{-7}$ & $7.0\times 10^{-8}$ \\
0.4 & $5.5\times 10^{-8}$ & $3.0\times 10^{-8}$ \\
\end{tabular}
\label{t1}
\end{center}
\end{table}

\end{document}